\documentclass[aps,prl,twocolumn,groupedaddress]{revtex4}
\usepackage{graphicx}

\usepackage{dcolumn}

\usepackage{bm}
\usepackage{color}
\usepackage[colorlinks]{hyperref}
\input{epsf}

\begin{document}


\title{Spin characterization and control over the regime of radiation-induced zero-resistance
states}

\author{R. G. Mani}
\email{mani@deas.harvard.edu} \affiliation {Harvard University,
Gordon McKay Laboratory of Applied Science, 9 Oxford Street,
Cambridge, MA 02138, USA}
%
%
%
%
\date{\today}
\begin{abstract}
Over the regime of the radiation-induced zero-resistance states
and associated oscillatory magnetoresistance, we propose a low
magnetic field analog of quantum-Hall-limit techniques for the
electrical detection of electron spin- and nuclear magnetic-
resonance, dynamical nuclear polarization via electron spin
resonance, and electrical characterization of the nuclear spin
polarization via the Overhauser shift. In addition, beats observed
in the radiation-induced oscillatory-magnetoresistance are
developed into a method to measure and control the zero-field spin
splitting due to the Bychkov-Rashba and bulk inversion asymmetry
terms in the high mobility GaAs/AlGaAs system.
\end{abstract}
%
%
\maketitle The interest in spintronics and spin-based
semiconductor quantum computing has increased the regard for the
spin degree of freedom, and especially semiconductor systems that
allow for the control of spin.[1-12] Low Dimensional Electronic
Systems[LDES] have consequently drawn special attention because,
at low temperatures, $T$, they include properties, which make
possible relatively simple electrical detection- and
radio-frequency ($rf$) / microwave control- of nuclear- and
electronic- spins. For example, microwave-induced Electron Spin
Resonance (ESR) can be resistively detected in the quantum Hall
regime, the ESR can be utilized to build up nuclear polarization
via the flip-flop interaction, and the nuclear spin state can be
subsequently characterized, also using an electrical measurement,
by examining the back action of the nuclear magnetic field on the
ESR.[13-19] Such electrical characterization techniques are
valuable because they can potentially help to characterize the
spin state over microscopic length scales.

Our studies have examined the above-mentioned approach towards the
initialization, control, and readout of the nuclear spin
polarization in spin domains within the LDES, with a view towards
spintronic and quantum computing applications. The goals have been
to develop the capability to measure and control a spin domain, to
scale down its size to reduce the number of spins per domain, and
to finally realize the capacity to simultaneously handle, i.e.,
measure and control, a multiplicity of domains on a single
chip.[16-19]

Here, we propose the possibility of extending these quantum Hall
spin measurement and control techniques into a low magnetic field
limit, by building upon the recent observation of novel
zero-resistance states,[20] which can be artificially induced and
easily tuned by microwave excitation of the 2DES, at low magnetic
fields, $B$ $\leq$ 0.5 T.[20-50] This new radiation-induced
physical effect, which brings quantum-Hall-effect-like vanishing
resistance states and activated transport characteristics to
extremely low magnetic fields,[20,21] suggests possible future
resistive spin detection and $rf$/microwave spin - manipulation
techniques in the vicinity of zero magnetic field. An attractive
feature of operating in the vicinity of zero-magnetic-field is
that it is likely to increase the scope of possible future
applications by simplifying device implementation.

Thus, we suggest the electrical detection of ESR in the domain of
the microwave induced zero-resistance states, with the aim of
utilizing the ESR to build up nuclear spin polarization in a weak
magnetic field limit. A polarized nuclear spin system might
subsequently be characterized at low $B$ through the detection of
the Overhauser shift of the electrically detected ESR in the
regime of the radiation induced zero-resistance states.[51] This
approach might also be utilized to measure the spin relaxation and
coherence times in the vicinity of null magnetic field.

Additionally, a lineshape study of beats in the radiation induced
oscillatory magnetoresistance is applied to extract the electronic
spin-splitting due to spin-orbit effects in the vicinity of
zero-magnetic field.[27] A suggestion is also made for the
measurement, manipulation, and control of the Zero-field Spin
Splitting (ZFSS) using this technique.

The paper is organized as follows: Section II presents background
and related work on electrical detection and the rf/microwave spin
techniques in the quantum Hall regime. Section III(A) illustrates
experimental results, which demonstrate the functionality of the
approach described in section II. Section III(B) exhibits the
relevant characteristics of the radiation-induced zero-resistance
states observed in the low field limit. Section III(C) examines
the suggested approach for characterizing the zero-field spin
splitting in high mobility 2D electron systems using the
radiation-induced resistance oscillations. Finally, section IV
extends, by analogy, the quantum Hall spin detection techniques
into the regime of the radiation-induced zero-resistance states,
and sketches the control, using measurement feedback, of the
zero-field spin splitting in the GaAs/AlGaAs system.
\begin{figure}[t]
\begin{center}
\includegraphics*[scale = 0.25,angle=0,keepaspectratio=true,width=3.0in]{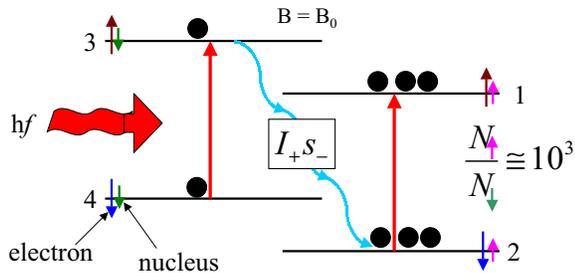}
\caption{(color online) The nuclear spin polarization, at a given
temperature, can be boosted by saturating the Electron Spin
Resonance (ESR) in a coupled electron-nucleus system, see ref.
51.}\label{1}
\end{center}
\end{figure}

\section{Background}
A two-dimensional electron system (2DES) at high magnetic fields
exhibits Landau quantization, and disorder leads to level
broadening, along with the localization of states between the
Landau subbands, in the low $T$ limit.[13] When the Fermi level in
the 2DES is pinned amongst the localized states, a quantized Hall
effect is manifested by zero-resistance states, i.e., $R_{xx}$
$\rightarrow$ 0, and quantized plateaus in the Hall resistance,
$R_{xy}$. Here, $R_{xx}$ $\rightarrow$ 0 is the asymptotic
behavior in the $T \rightarrow 0$ limit. In practice, at liquid
helium temperatures, a finite resistance is observed, and $R_{xx}$
follows an activation law, i.e., $R_{xx} \sim exp (-T_{0}/T)$,
where $T_{0}$ is an activation energy.[13]

A non-vanishing electronic $g$-factor removes the spin degeneracy.
When the Fermi level is pinned in the localized states between
spin split subbands, the system exhibits, once again, activated
transport leading into zero-resistance states. Then, the
conditions become favorable for the measurement and control, with
spatial resolution, of nuclear and electronic spins in such
systems. In particular, Electron Spin Resonance(ESR) induced, for
example, by microwave excitation, becomes observable in the
electrical response.[14,17-19,51] Further, under steady state ESR,
the decay of spin-excited electrons leads to the spin polarization
of nuclei via the flip-flop, electron-nuclear hyperfine
interaction. The magnetic field arising from the spin polarized
nuclei then provides a back action on the electrons, leading to an
Overhauser shift in the electrically detected ESR, which can then
serve to characterize the magnetic state of the nuclear spin
system.
\begin{figure}[t]
\begin{center}
\includegraphics*[scale = 0.25,angle=0,keepaspectratio=true,width=3.0in]{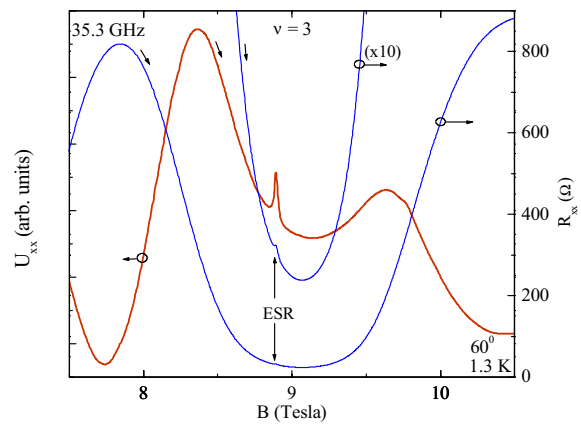}
\caption{(color online) The figure shows the diagonal ($R_{xx}$)
resistance of a GaAs/AlGaAs device in the vicinity of filling
factor $\nu$ = 3, at T = 1.3 K and a tilt angle of $60^{0}$. The
microwave induced voltage $U_{xx}$ shows Electron Spin Resonance
(ESR) under photoexcitation at $f$ = 35.3 GHz. }\label{2}
\end{center}
\end{figure}

The hyperfine (spin-spin) interaction between an electron, $S$,
and a nucleus, $I$, $A(I\cdot S)$ = $A((I_{+}S_{-} + I_{-}S_{+})/2
+ I_{z}S_{z})$,[14,51] implies a four - level system for
spin-$1/2$ particles (see Fig. 1). The saturation of the electron
spin resonance by the application of intense microwave radiation
tends to equalize the electronic population of the levels, while
the phonon-assisted 'flip-flop'  exchange of spin between
electrons and nuclei through the off-diagonal term provides a
mechanism for realizing a steady state, see Fig. 1, which includes
a large population difference between the nuclear spin
states.[17-19,51] As the effective Boltzmann factor describing the
steady state nuclear spin polarization comes to be determined by
the electronic spin-flip energy (Fig. 1), it is, in principle,
possible to realize a large nuclear spin polarization using this
technique at a relatively high temperature.[51] A build up of the
nuclear spin polarization by ESR also generates a nuclear magnetic
field, $B_{N}$, via the nuclear magnetic moment, which modifies
the spin resonance condition for electrons. Thus, there is a
characteristic B-field shift of the ESR with the spin polarization
of nuclei, as mentioned previously, that is proportional to the
nuclear magnetic field, $B_{N}$. This Overhauser shift makes it
possible to characterize the nuclear spin state through a
measurement of the electrically detected ESR. In addition, dynamic
nuclear polarization implies that the electrical resistance
becomes sensitive to the nuclear spin state, suggesting also the
possibility of nuclear magnetic resonance detection through a
resistance measurement.[14-19]
\begin{figure}[t]
\begin{center}
\includegraphics*[scale = 0.25,angle=0,keepaspectratio=true,width=3.0in]{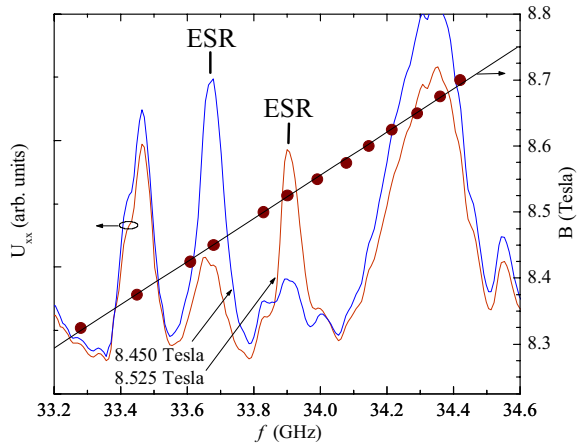}
\caption{(color online) Measurements of Electron Spin Resonance
(ESR), at fixed magnetic fields, as a function of the microwave
frequency, in a GaAs/AlGaAs device. The signature of ESR is the
strong $B$-sensitivity of the $U_{xx}$ signal over narrow
$f$-intervals. }\label{3}
\end{center}
\end{figure}

Thus, such physical phenomena can, in principle, be utilized to
realize discrete, independently controlled, and separately
measured nuclear spin domains. In particular, nuclear spin
initialization to a logic $'1'$ state can be achieved at a
relatively high temperature by applying dynamic nuclear
polarization. A local operation on a single domain can be
implemented by introducing or removing electrons into the nuclear
neighborhood using a voltage controlled 'gate,' as the specimen is
irradiated with a global microwave field. A $'0'$ state would then
follow from a $'1'$ state upon spin-rotation with a
radio-frequency $\pi$-pulse. A superposition of $'1'$ and $'0'$
states might be realized by initializing a domain and then
subjecting it to resonant radio frequency $\pi/2$ pulse. Readout
of the nuclear spin state is accomplished by detecting the shift
in the electrically detected ESR due to the nuclear magnetic
field. Here, the $'1'$ and $'0'$ states will exhibit opposite
Overhauser shifts, while state superposition might be reflected as
an oscillatory Overhauser shift.[16-19] According to the
theoretical studies of Taylor and co-workers,[52] an ensemble of
nuclei, i.e., a nuclear spin domain, can serve as a repository or
storage media for the quantum information associated with a mobile
electronic qubit, and it is possible, in principle, to transfer
quantum information in both directions, between the mobile
electronic qubit and the nuclear spin domain with high fidelity.
The operations described above can serve, for example, to prepare
the spin domains for such a function.
\begin{figure}[t]
\begin{center}
\includegraphics*[scale = 0.25,angle=0,keepaspectratio=true,width=3.0in]{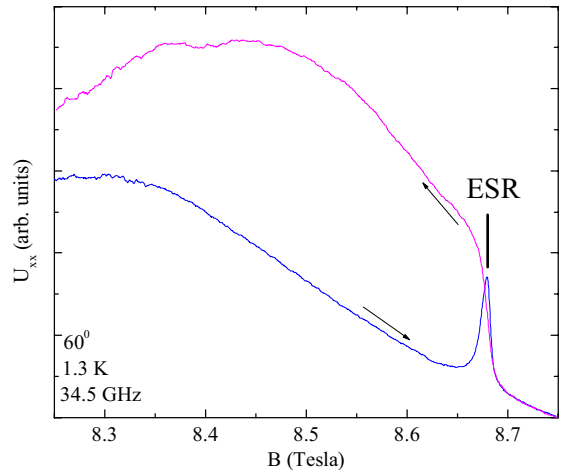}
\caption{(color online) Sweep-direction dependent hysteresis in
the ESR signal in the vicinity of filling factor $\nu$ = 3 at a
tilt angle of $60^{0}$ indicates Dynamic Nuclear Polarization
(DNP) and an Overhauser shift.}\label{4}
\end{center}
\end{figure}
\section{Experiment}
\subsection{Spin measurement and control in the quantum Hall
regime}
In this section, some concepts associated with the
above-mentioned approach are illustrated through experiment. For
this purpose, measurements were carried out on GaAs/AlGaAs
specimens, which were mounted inside a waveguide, in  a low
temperature cryostat including a superconducting magnet.  The
samples were typically irradiated with amplitude-modulated
microwaves over the frequency range $27 \leq f \leq 60$ GHz, as a
double lock-in technique was employed to extract the microwave
induced signal, $U_{xx}$, that exhibits ESR. For NMR measurements,
the specimens were subjected to simultaneous microwave and
\textit{rf} excitation, with the \textit{rf} spanning the range
$30 \leq f \leq 100$ MHz. [16-19]

Figure 2 shows the typical transport response of a 2DES device
under microwave excitation at 35.3 GHz. In this swept-$B$,
fixed-$f$ experiment, the microwave excitation helps to realize a
resonance condition, when the photon energy, $hf$, matches the
level spacing of the electron spin-split bands, and this leads to
a resonant heating of the electron system, which is manifested as
a detectable change in the electrical resistance. As illustrated
in Fig. 2, the microwave induced signal $U_{xx}$, exhibits such
ESR in the vicinity of $B$ $\approx$ 9 Tesla.

Figure 3 exhibits electrically detected ESR in a complementary
fixed-$B$, swept-$f$ experiment. The fixed-$B$ experimental data
shown in Fig. 3 indicate non-monotonic, $B$-insensitive,
microwave-induced signals $U_{xx}$, which are attributed to
$f$-dependent radiation intensities at the sample. In addition,
there is a strong $B$-sensitive response vs $f$, which is the ESR
signal. In order to characterize the resonance, the lineshape of
the ESR signal was fit with Gaussian and Lorentzian lineshapes.
Such fits indicated an inhomogeneous broadening contribution to
$T_{2}^{*}$ for electrons.
\begin{figure}[t]
\begin{center}
\includegraphics*[scale = 0.25,angle=0,keepaspectratio=true,width=3.0in]{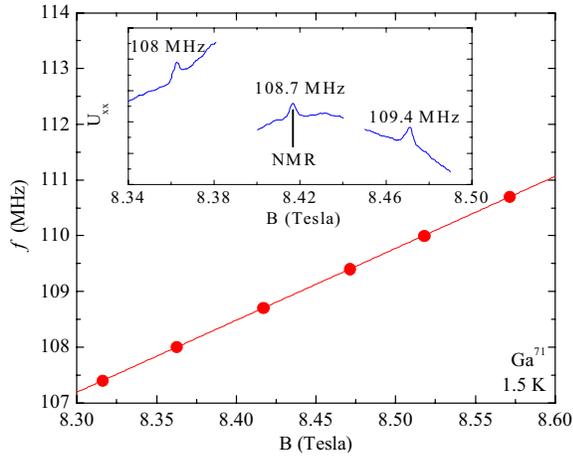}
\caption{(color online) (inset) Electrical detection of Nuclear
Magnetic Resonance of $Ga^{71}$ nuclei in a GaAs/AlGaAs device
with radio frequency excitation as indicated. The main panel shows
the linear shift of the NMR frequency with increasing magnetic
field. Measurements were carried out with microwave excitation at
34.5 GHz, at a tilt angle of $60^{0}$.}\label{5}
\end{center}
\end{figure}

A role for dynamic nuclear polarization in such experiments can be
motivated by exhibiting hysteretic transport in, for example,
fixed-$f$, swept-$B$ measurements, of the type shown in Fig. 4.
Here, in Fig. 4, a broad linewidth observed in the down-sweep
measurement is a manifestation of the Overhauser shift since a
nuclear magnetic field $B_{N}$ compensates for the reduced applied
magnetic field, and helps to maintain the ESR condition even below
the $B$ where ESR would normally occur.

The role for nuclear polarization in this line broadening can be
confirmed by examining the sensitivity of the electrical
resistance to nuclear magnetic resonance (NMR). For this purpose,
microwave and \textit{rf} excitation were simultaneously applied
to the specimen, with the \textit{rf} chosen to coincide with the
nuclear magnetic resonance frequency of the host nuclei, i.e., Ga
or As. The observation of electrically detected NMR, illustrated
in Fig. 5, confirm that enhanced nuclear polarization is
responsible for the hysteretic transport observed in Fig. 4.
\begin{figure}[t]
\begin{center}
\includegraphics*[scale = 0.25,angle=0,keepaspectratio=true,width=3.0in]{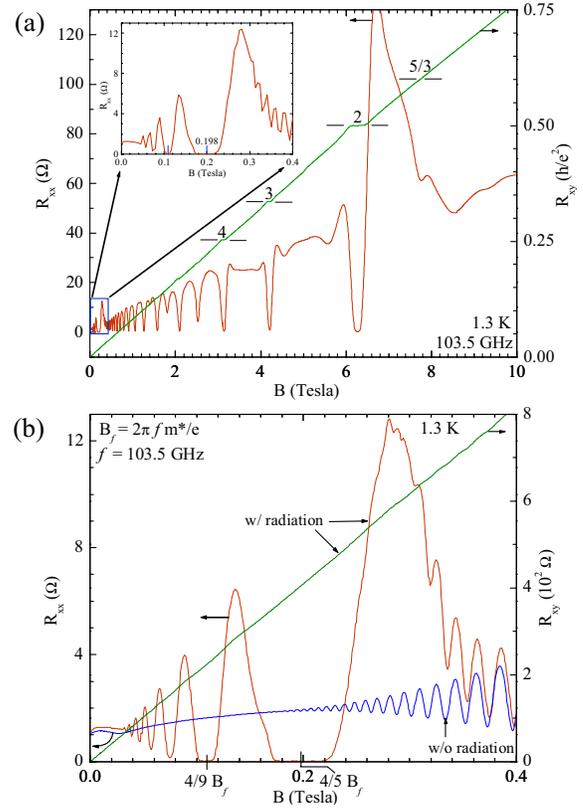}
\caption{(color online) (a) Measurements of the Hall ($R_{xy}$)
and diagonal ($R_{xx}$) resistances in a high mobility GaAs/AlGaAs
heterostructure under excitation at 103.5 GHz. Quantum Hall
effects occur at high $B$ as $R_{xx}$ $\rightarrow$ 0. (inset) An
expanded view shows $R_{xx}$ $\rightarrow$ 0 in the vicinity of
0.2 Tesla. (b) Data over low $B$ obtained both with (w/) and
without (w/o) radiation at 103.5 GHz. Note the vanishing
resistance under photoexcitation in the vicinity of $ (4/5) B_{f}$
and $(4/9) B_{f}$, that goes together with a linear Hall
effect.}\label{6}
\end{center}
\end{figure}

\subsection{Novel radiation-induced zero-resistance states}
The spin manipulation and measurement techniques described in the
previous section operate in a quantum Hall regime characterized by
activated transport and vanishing resistance, typically at high
magnetic fields. Yet, the high field requirement is likely to
complicate real-world applications. The question then arises
whether there is a possibility of realizing and implementing a
similar approach in a low-$B$ limit. Basically, from the
perspective of the experimentalist, this seems to require bringing
the essential quantum Hall features, vanishing resistance and
activated transport, to the vicinity of zero magnetic field. We
describe new phenomena below, which help to bring such quantum
Hall features to the desired low magnetic field. The approach
includes the added benefit of simple tunability of the
zero-resistance states.

\begin{figure}[t]
\begin{center}
\includegraphics*[scale = 0.25,angle=0,keepaspectratio=true,width=3.0in]{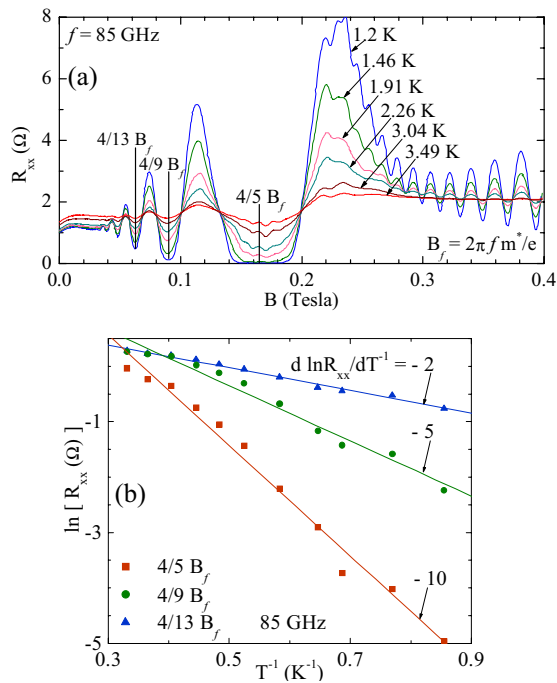}
\caption{(color online)(a) The temperature dependence of $R_{xx}$
at 85 GHz under constant-intensity microwave photoexcitation. The
radiation-induced resistance minima become deeper at lower
temperatures. (b) A plot of the logarithm of the resistance vs.
the inverse temperature at $B = (4/5) B_{f}$, $(4/9) B_{f}$, and
$(4/13) B_{f}$ suggests activated transport. The slope of these
curves indicates the activation energy.}\label{7}
\end{center}
\end{figure}
Briefly, experiments indicate that ultra high mobility GaAs/AlGaAs
heterostructures including a 2DES can exhibit novel
quantum-Hall-like vanishing resistance states, without Hall
resistance quantization, at low temperatures, $T$, and low
magnetic fields, $B$, when the specimen is subjected to
Electro-Magnetic (EM) wave excitation. Zero-resistance-state occur
about, for example, $B = (4/5) B_{f}$ and $B = (4/9) B_{f}$, of
the characteristic field $B_{f} = 2\pi f m^{*}/e$, where $m^{*}$
is the electron mass, $e$ is electron charge, and $f$ is the
EM-wave frequency, as the resistance-minima follow the series $B =
[4/(4j+1)] B_{f}$ with $j = 1,2,3,$... Temperature-dependent
measurements on the observed resistance-minima also indicate
activated transport.[20] Simply put, these novel zero-resistance
states exhibit the desired quantum Hall characteristics so far as
our objectives are concerned.

For the associated experiments, specimens were fabricated from
ultra high mobility GaAs/AlGaAs heterostructures exhibiting a
mobility $\mu (1.5 K)$  up to $1.5 \times 10^{7}$ $cm^{2}/Vs$.
Typically, the sample was once again mounted inside a waveguide,
immersed in pumped liquid Helium, and irradiated with microwaves,
this time over the range $27 \leq f \leq 120$ GHz.[20]

Fig. 6 (a) shows the diagonal $(R_{xx}$) and Hall ($R_{xy}$)
resistances measured in the four-terminal configuration. Here,
$R_{xx}$ and $R_{xy}$ exhibit the usual quantum Hall behavior for
$B$ $\geq$ 0.4 Tesla even under microwave excitation at $f$ =
103.5 GHz. In contrast, at $B$ $<$ 0.4 Tesla, see inset, Fig.
6(a), a radiation induced signal occurs and, remarkably, the
resistance vanishes over a broad B-interval about $B$ = 0.198
Tesla.
\begin{figure}[t]
\begin{center}
\includegraphics*[scale = 0.25,angle=0,keepaspectratio=true,width=3.0in]{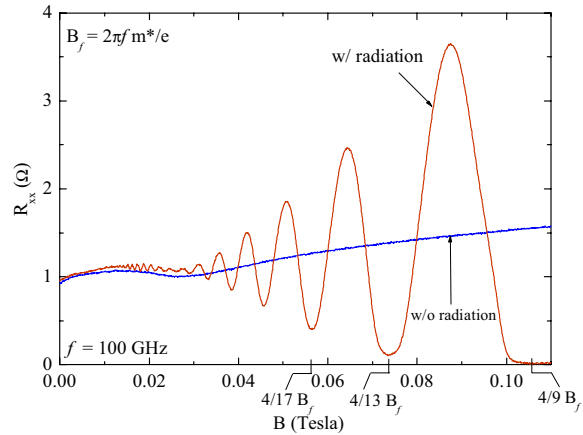}
\caption{(color online) The figure illustrates $R_{xx}$ in a high
mobility GaAs/AlGaAs heterostructure device with (w/) and without
(w/o) microwave excitation. A beat is observable about $B$
$\approx$ 0.024 Tesla, which is associated with a zero-field spin
splitting in this system.}\label{8}
\end{center}
\end{figure}

Further high-resolution measurements are shown in Fig. 6 (b).
Without EM-excitation, $R_{xx}$ exhibits Shubnikov-deHaas
oscillations for $B$ $\geq$ 0.2 Tesla. The application of
radiation induces additional oscillations, and, at the deepest
minima, $R_{xx}$ falls well below the resistance measured without
radiation, as it vanishes around $(4/5) B_{f}$ and $(4/9) B_{f}$.
Although these resistance minima saturate at zero-resistance as in
the quantum Hall regime, $R_{xy}$ does not exhibit plateaus over
the same $B$-interval. We have found that $R_{xx}$ minima occur
$B_{min}^{-1}/\delta$ = $[4/(4j+1)]^{-1}$ with $j =
1,2,3$...[20,48] As $B_{f}$ depends on $f$, the periodicity of
these radiation induced magnetoresistance oscillations changed
with $f$, and a given resistance minimum moved to higher B with
increasing $f$.

The temperature variation of $R_{xx}$ at 85 GHz, shown in Fig.
7(a), displays both the strong $T$-dependence of $R_{xx}$ and the
low-$T$ requirement for observation of the zero-resistance-states.
The $T$-variation of $R_{xx}$ at the deepest minima demonstrate
activated transport, i.e., $R_{xx}$ $\sim$ $exp(-T_{0}/T)$, see
Fig. 7(b), and the activation energy exceeds the Landau level
spacing.[20,21]

These fundamental features, vanishing resistance and activated
transport, are strongly reminiscent of the quantum Hall
situation.[13] Indeed, since these two key features are the active
ingredients, from the experimental perspective, for the
functioning of the approach described in section II and III(A),
there appears to be the possibility of utilizing these vanishing
resistance states in lieu of the quantum Hall zero-resistance
states, towards developing a low field analog of the approach
suggested in section III(A).

It is true that there are measurable differences in the electrical
transport between the regime of the radiation induced
zero-resistance states and the regime of the quantum Hall effect,
such as, e.g., the absence of Hall resistance quantization in the
case of the microwave induced effect (Fig. 6(b)). Yet, since the
Hall resistance is not actively used in the experimental
measurements of spin resonance as in Figs. 2 - 5, we expect that
Hall resistance quantization, or lack thereof, will not modify the
functioning of the experimental technique. It appears that a key
ingredient for the operation of the techniques of section III(A)
is the suppression of the diagonal resistance, which originates
from the residual scattering in the 2DES. Once this contribution
has been suppressed, the resistance tends to reflect a special
sensitivity, due to "activated transport," to the microwave
induced ESR, and this is the physical feature that makes possible
electrical detection of spin resonance.[14] From this perspective,
it appears that the radiation induced zero-resistance states ought
to serve as an adequate replacement for the quantum Hall
resistance states.

\subsection{Determination of the zero-field spin splitting from the
radiation-induced oscillatory resistance} In the previous
section,we demonstrated novel radiation induced zero-resistance
states in the GaAs/AlGaAs system and reasoned that they could
serve in lieu of quantum Hall zero-resistance states at low-$B$ in
the development of electrical spin detection techniques for low
magnetic field applications. Before further exploring such a
possibility in the next section, we show here that beats observed
in the low field radiation-induced resistance oscillations might
serve as a sensitive probe of the zero-field spin splitting
originating from the Bychkov-Rashba term and Bulk Inversion
Asymmetry term in the high mobility 2DES.[53,54]

It is well known that mobile 2D electrons can experience, in their
rest frame, an effective magnetic field that develops from a
normal electric field at the semiconductor heterojunction
interface due to the so-called Bychkov-Rashba effect.[53] As this
magnetic field can, in principle, be controlled by an electrical
gate, it has been utilized in the design of novel spin based
devices such as the spin transistor.[2] In the 2DES realized in
wider gap GaAs/AlGaAs, a Bulk Inversion Asymmetry (BIA)
contribution to the ZFSS, can also provide a "Zeeman magnetic
field" as $B$ $\rightarrow$ 0.[54-56] Although theory has
suggested that the BIA term is stronger than the Rashba term in
the GaAs/AlGaAs heterostructure 2DES,[57] ZFSS in $n$-type
GaAs/AlGaAs is not easily characterized because its signature is
often difficult to detect using available methods.

Typical investigations of ZFSS in the 2DES look for beats in the
Shubnikov-de Haas (SdH) oscillations that originate from
dissimilar Fermi surfaces for spin-split bands.[58-60] Yet, it is
known that other mechanisms can, in principle, provide similar
experimental signatures.[59-61] Thus, a need has existed for new
methods of investigating the spin-orbit interaction in the 2DES
characterized by a small zero-field spin splitting, to supplement
the Electron Spin Resonance, SdH, Raman scattering, and
weak-localization based approaches.[14,59,62,63]

In this light, the approach outlined below is simpler and provides
improved sensitivity because the ZFSS ($\approx$ 20 $\mu$eV) is
determined through a comparison of the spin splitting with an
easily tunable, small energy scale set by the photon energy
($\approx$ 200 $\mu$eV), unlike the SdH approach which relates the
ZFSS ($\approx$ 20 $\mu$eV)  to differences between two (spin
split) Fermi surfaces with energy of order $E_{F}$ $\approx$ 10
meV in GaAs/AlGaAs.

Fig. 8 illustrates the magnetoresistance $R_{xx}$ observed with
(w/) and without (w/o) microwave excitation, in a high mobility
condition. The figure shows that, with radiation, there occur
oscillations in $R_{xx}$ as described in the previous section.
Additionally, these oscillations show a non-monotonic variation in
the amplitude, i.e., a beat, at low $B$ (see Fig. 8), as they
extend to lower-$B$. A study of the such transport data obtained
at different $f$ revealed that, remarkably, the beat remains at a
fixed $B$ with a change in the microwave frequency. Thus, a
lineshape study was carried out in order to characterize this
beating effect. Over a narrow $B$-window above the beat, the
oscillatory data could be fit with a single exponentially damped
sinusoid: $R_{xx}^{osc}$ = $A' exp(-\lambda /B)sin(2\pi F/B -
\pi)$, where $A'$ is the amplitude, $\lambda$ is the damping
factor, and $F$ is the resistance oscillation frequency. Yet, a
lineshape that included  a superposition of two such waveforms
proved unsatisfactory in modelling the data, when the data
exhibited beats.

Thus, a candidate waveform $R_{xx}^{osc} = A exp(-\lambda /B)
sin(2 \pi F/B - \pi) [1 + cos(2 \pi \Delta F/B)]$, which can
realize beats without phase reversal, was applied in the analysis.
Representative data and fit, see Fig 9, show that this lineshape
describes the data quite well. A summary of the fit parameters,
$F$, $\Delta F$, and $\lambda$, is presented in Fig. 10. A
noteworthy feature here is that the beat frequency, $\Delta F$
$\approx$ 12.3 mTesla, is independent of $f$ (see Fig. 10(b)).

This applied lineshape can be understood by invoking four distinct
transitions between Landau subbands near the Fermi level, as in
the inset of Fig. 9. Here, the spin-orbit interaction helps to
remove the spin degeneracy of Landau levels as $B$ $\rightarrow$
0. If the oscillations originating from these terms have the same
amplitude and share the same $\lambda$, then one expects a
superposition of four terms: $R_{xx}^{osc}$ = $A' exp(-\lambda /B)
[sin(2 \pi F/B - \pi) + sin(2 \pi F/B - \pi) + sin(2 \pi [F-\Delta
F]/B - \pi) + sin(2\pi [F + \Delta F]/B - \pi)] = A exp(-\lambda
/B) sin(2 \pi F/B - \pi) [1 + cos(2 \pi \Delta F/B)]$, which
constitutes the lineshape that has been utilized to fit the data.
\begin{figure}[t]
\begin{center}
\includegraphics*[scale = 0.25,angle=0,keepaspectratio=true,width=3.0in]{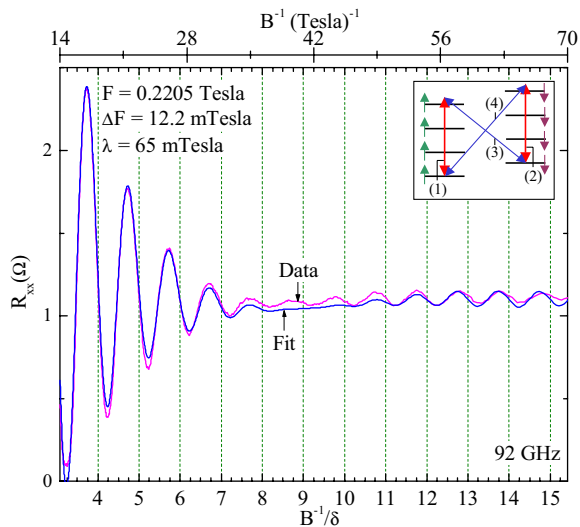}
\caption{(color online) $R_{xx}$ oscillations are shown at 92 GHz,
along with a lineshape fit (see text) vs. $B^{-1}/\delta$, where
$\delta$ = $F^{-1}$. The lineshape fit includes the contributions
shown in the inset. Here, as $B$ $\rightarrow$ 0, the spin-up and
spin-down Landau levels are shifted in energy due to a zero-field
spin splitting.}\label{9}
\end{center}
\end{figure}

Thus, beats observed in the radiation-induced resistance
oscillations can be understood as a consequence of a zero-field
spin splitting, due to the spin-orbit interaction.[2,53-60,62-66]
One might relate $\Delta F$ to the ZFSS by identifying the
radiation-frequency change, $\delta f$, that will produce a
similar change in $F$, i.e., $\delta f$ = $\Delta F/[dF/df ]$,
where $\Delta F$ is the beat frequency, and $dF/df$ is the rate of
change of $F$ with the radiation frequency (see Fig. 10(a)). Then,
the ZFSS corresponds to $\delta f$ = 5.15 GHz or $E_{S}$($B$ = 0)
= $h \delta f$ = 21 $\mu$eV. In comparison, theory suggests that,
in the GaAs/AlGaAs heterostructure 2DES, the upper bound for the
total ZFSS is $\approx$ 80 $\mu$eV.[57]

A ESR study in the quantum Hall regime, as described in section
III (A), showed that the ESR field, $B_{ESR}$, varied as $dB_{ESR}
/d f$ = 0.184 Tesla/GHz in the vicinity of filling factor $\nu$ =
1, which suggests that the effective Zeeman magnetic field is
approximately $B_{Z}$ = $[dB_{ESR} /d f ] \delta f$ = 0.95 Tesla.
Here, this $B_{Z}$ is identified as the magnetic field that
appears in the rest frame of the electron, in the absence of an
applied magnetic field, due to the spin-orbit effects mentioned
above.

\section{Extension of quantum Hall spin detection techniques to the
regime of the radiation-induced zero-resistance states, and
measurement and control of zero-field spin splitting} As reasoned
above, electronic systems which exhibit vanishing resistance
states and activated transport at high magnetic fields, include
special features that make possible electrical detection of ESR
and NMR. In section II, the essential physics behind this approach
in the quantum Hall regime was reviewed. In section, III(A),
experimental data were exhibited, which illustrated the viability
of this approach in the quantum Hall regime. Although the concepts
and techniques functioned as expected in the quantum Hall regime,
the desirability of realizing a similar scheme in the low magnetic
field limit was highlighted as part of the effort towards
simplifying future device implementation. In section III(B), we
demonstrated novel radiation induced zero-resistance states at low
magnetic fields that included both the activated transport- and
vanishing resistance states- characteristics of quantum Hall
systems. In section III(C), we also suggested that beats in these
novel radiation induced resistance oscillations might serve to
directly measure the zero field spin splitting and the Zeeman
magnetic field that appears in the rest frame of the electron due
to the spin-orbit interaction.

Here, we outline, by analogy, an experimental scenario for
electrical spin detection and dynamic nuclear polarization in the
low $B$ limit, over the regime of the radiation-induced
zero-resistance states: A high mobility 2DES might be irradiated
with microwaves, at a frequency $f_{0}$, in order to induce
magnetoresistance oscillations, as in section III(B). For the
appropriate radiation intensity, a zero-resistance state would be
manifested in the vicinity of $B$ = (4/5) $B_{f}$, where $B_{f} =
2\pi f_{0} m^{*}/e$. Upon realizing the desired zero-resistance
condition, a second radiation field at frequency $f_{1}$ is
applied to the specimen in order to induce spin flip transitions
or ESR in the vicinity of $(4/5) B_{f}$. The frequency of this
radiation, $f_{1}$, should be selected to match the total spin
splitting. Due to the strong temperature sensitivity of the
resistance at the radiation-induced zero-resistance states
resulting from activation transport characteristics, see Fig 7
(b), ESR induced by the radiation field ought to be manifested as
a change in the electrical resistance, just as in the quantum Hall
situation (Fig. 2). One expects this behavior because,
phenomenologically, the radiation at $f_{0}$ helps to suppress the
residual scattering in the 2DES, and this should lead to an
enhanced sensitivity to the ESR induced by $f_{1}$, which might be
viewed as a new resonant scattering channel.

\begin{figure}[t]
\begin{center}
\includegraphics*[scale = 0.25,angle=0,keepaspectratio=true,width=3.0in]{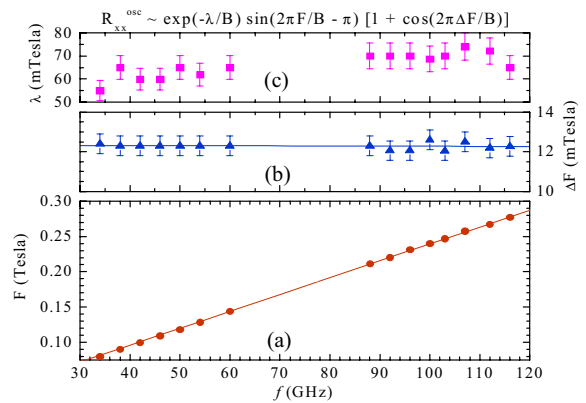}
\caption{(color online)) Fit parameters obtained from the
lineshape study of the radiation-induced $R_{xx}$ oscillations.
(a) The resistance oscillation frequency, $F$, increases linearly
with the radiation frequency. (b) The beat frequency, $\Delta F$,
is independent of $f$, and $\Delta F$ $\approx$ 12.3 mTesla. (c)
The damping parameter $\lambda$ is also independent of
$f$.}\label{10}
\end{center}
\end{figure}

A steady state ESR, perhaps in the presence of an oriented
current, might then be applied to dynamically polarize nuclei via
the flip-flop interaction. This could be manifested as a
broadening of the ESR line in analogy to Fig. 4. One might now
track the time evolution of the ESR line to measure the nuclear
spin relaxation rate at low $B$, and study its sensitivity to
magnitude of the diagonal resistance of the specimen, which can be
tuned with the radiation intensity at $f_{0}$. Finally, one might
apply an $rf$-field at a frequency $f_{2}$ in order to induce- and
electrically detect- nuclear magnetic resonance, as in Fig. 5.

A goal would be to apply dynamic nuclear polarization and obtain
polarized nuclear spin domains in the low magnetic field limit.
Once polarized nuclear spin domains have been realized using the
approach outlined above, $rf$-pulses might be applied to rotate
the spin domain. Alternatively, one might examine the effect of
magnetic field reversal with respect to the domain polarization
axis, through such experiments.

The zero-field spin splitting measurement scheme of section III(C)
might also be applied towards the characterization and control of
the electronic spin splitting due to the Bychkov-Rashba and
Dresselhaus terms in GaAs/AlGaAs.[53,54,64] According to Tarasenko
and Averkiev,[64] beats should be observable when either the
Bychkov-Rashba or the Dresselhaus terms are manifested in the
specimen, and the beats should disappear when the two
contributions are equal in magnitude.  Thus, in a high mobility
quantum well with a negligible electric field along the growth
direction, one might first investigate the beats originating from
the Dresselhaus effect. Then, a top-gate might be utilized to
change the electric field along the growth direction and
change/introduce the Bychkov-Rashba term.[64]

Perhaps, the top-gate can also be applied to change the electron
spin resonance condition at a fixed magnetic field. Then, if one
wishes to preferentially polarize the nuclear spin on one domain,
while leaving the nuclear spin of other domains unchanged, the
Bychkov-Rashba effect can be utilized to bring into- or take out
of- the electron spin resonance condition the domain of interest,
with the help of the top-gate, since the gate influence will be
local only to the addressed domain.

\section{Summary}
The manipulation, measurement, and control of electronic and
nuclear spin have been important themes in the spintronics
paradigm,[1] which aims to realize novel devices that utilize also
the spin degree of freedom.[65,66] Here, we have proposed spin
measurement, manipulation, and control techniques, in a low
magnetic field limit, in the high mobility GaAs/AlGaAs system. In
particular, we have suggested the possibility of operating on
nuclear spins via electrons and electrically detecting the
results, at low magnetic fields, over the regime of the recently
discovered radiation-induced zero-resistance states in the
GaAs/AlGaAs system. As discussed, one might try to extend by
analogy the experimentally demonstrated electrical-detection,
rf/microwave manipulation techniques from the quantum Hall regime,
into the low field limit, based on the observed phenomenological
similarity of transport in the high field quantum Hall- and the
low field radiation induced zero-resistance- regimes. Success in
this direction might lead to discrete, independently controlled,
and separately measured nuclear spin memory cells that function at
low magnetic fields, and might potentially serve as a long-term
repository for the quantum information associated with a mobile
electronic qubit.[52]

We have also identified a possible new method for characterizing
the ZFSS. In particular, observed beats in the radiation-induced
magnetoresistance oscillations have been suggested as a tool for
characterizing the zero-field spin splitting originating from the
Bychkov-Rashba effect and BIA in the GaAs/AlGaAs system.[53,54]
Characterization and control of the zero-field spin splitting in
GaAs/AlGaAs might advance this widely available material, which
provides the highest mobility 2DES, as a potential host for the
realization of the spin based devices at low-$B$.

 \vspace{0cm}
\begin{acknowledgments}
We acknowledge stimulating discussions with S. Saykin, M. Dobers,
J. H. Smet, K. von Klitzing, V. Narayanamurti, W. B. Johnson, V.
Privman, and D. Mozyrsky. High mobility material was kindly
provided by V. Umansky.
\end{acknowledgments}


\begin{thebibliography}{66}

\bibitem{1}  S. A. Wolf et al., "Spintronics: A spin-based electronics vision for the
future," \textit{Science }\textbf{294}, 1488-1495 (2001).

\bibitem{2}  S. Datta and B. Das, "Electronic analog of the electro-optic
modulator," \textit{Appl. Phys. Lett.} \textbf{56}, 665-667
(1990).

\bibitem{3}  D. P. DiVincenzo, "Quantum Computation," \textit{Science} \textbf{270}, 255
(1995).

\bibitem{4}  C. H. Bennett, "Quantum Information and Computation," \textit{Physics
Today} 25-30 (Oct.1995).

\bibitem{5} L. K. Grover, "The advantages of superposition",
Science \textbf{280}, 5361 (1998).

\bibitem{6}  B. E. Kane, "A Silicon-based Nuclear Spin Quantum
Computer," \textit{Nature }\textbf{393}, 133 (1998).

\bibitem{7}  D. Loss and D. P. DiVincenzo, "Quantum Computation with Quantum
Dots," \textit{Phys. Rev. A }\textbf{57}, 120 (1998).

\bibitem{8}  A. Imamoglu et al., "Quantum Information Processing
Using Quantum Dot Spins and Cavity-QED," \textit{Phys. Rev. Lett.}
\textbf{83}, 4204 (1999).

\bibitem{9}  I. L. Chuang, N. Gershenfeld, and M. Kubinec, "Experimental
implementation of fast quantum searching," \textit{Phys. Rev.
Lett. }\textbf{80}, 3408 (1998).

\bibitem{10}  V. Privman, I. D. Vagner, and G. Kventsel, "Quantum
Computation in Quantum-Hall Systems," \textit{Phys. Lett. A}
\textbf{239}, 141 (1998).

\bibitem{11} R. Vrijen et al., "Electron-spin
resonance transistors for quantum computing in silicon-germanium
heterostructures," \textit{Phys. Rev. A} \textbf{62}, 012306
(2000).

\bibitem{12} T. D. Ladd et al., "All silicon quantum computer," Phys. Rev. Lett.
\textbf{89}, 017901 (2002).

\bibitem{13} R. E. Prange and S. M. Girvin, (eds) \textit{The Quantum Hall
Effect}, 2nd Ed., New York: Springer-Verlag, 1990.

\bibitem{14} M. Dobers, K. von Klitzing, J. Schneider, G. Weimann, and K.
Ploog, "Electrical Detection of Nuclear Magnetic Resonance in
GaAs/AlGaAs Heterostructures," \textit{Phys. Rev. Lett.}
\textbf{61}, 1650 (1988).

\bibitem{15} T. Machida et al., "Spin polarization of fractional
quantum Hall edge channels studied by dynamic nuclear
polarization," \textit{Phys. Rev. B}\textbf{65}, 233304 (2002).

\bibitem{16} R. G. Mani, W. B. Johnson, V. Narayanamurti, V. Privman and Y.
H. Zhang, "Nuclear Spin Memory and Logic in Quantum Hall
Semiconductor Nanostructures for Quantum Computing Applications,"
\textit{Physica E }\textbf{12}, 152 (2002).

\bibitem{17} R. G. Mani, W. B. Johnson, and V. Narayanamurti, "Manipulation
and measurement of nuclear spin over the quantum Hall regime for
quantum information processing," \textit{Superlattices and
Microstructures} \textbf{32}, 261 (2002).

\bibitem{18} R. G. Mani, W. B. Johnson, and V. Narayanamurti,
"Initialization of a nuclear spin system over the quantum Hall
regime for quantum information processing," in \textit{Proc. of
15th Intl. Conf. on the Appl. of High Magnetic Fields in Semicond.
Physics,} Oxford, Aug. 2002, IOP Conf. Ser. No. \textbf{171}, eds.
A. R. Long and J. H. Davies (IOP, Bristol, 2003) 1.6.

\bibitem{19} R. G. Mani, W. B. Johnson, and V. Narayanamurti, "Nuclear spin
based quantum information processing at high magnetic fields,"
\textit{Nanotechnology} \textbf{14}, 515 (2003).

\bibitem{20} R. G. Mani, J. H. Smet, K. von Klitzing, V. Narayanamurti, W. B. Johnson, and V. Umansky, "Zero-resistance states induced by
electromagnetic-wave excitation in GaAs/AlGaAs heterostructures,"
\textit{Nature (London)} \textbf{420}, 646 (2002).

\bibitem{21} M. A. Zudov, R. R. Du, L. N. Pfeiffer, and K. W. West, "Evidence
for a new dissipationless effect in 2D electronic transport,"
\textit{Phys. Rev. Lett.} \textbf{90}, 046807 (2003).

\bibitem{22} R. Fitzgerald, "Microwaves induce vanishing resistance in the
two-dimensional electron system," \textit{Phys. Today }\textbf{56}
(4), 24-27 (2003).

\bibitem{23} R. G. Mani, J. H. Smet, K. von Klitzing, V. Narayanamurti, and V. Umansky,
"Single particle and collective response in the
magnetophotoresistance of a high mobility 2DES under microwave
excitation," \textit{Bull. Am. Phys. Soc.} \textbf{46}, p. 972
(2001).

\bibitem{24} M. A. Zudov, R. R. Du, J. A. Simmons, and J. L. Reno,
"Shubnikov-de Haas-like oscillations in millimeterwave
photoconductivity in a high-mobility two-dimensional electron
gas," \textit{Phys. Rev. B} \textbf{64}, 201311 (2001).

\bibitem{25} P. D. Ye et al., "Giant microwave photoresistance of
two-dimensional electron gas," \textit{Appl. Phys. Lett.
}\textbf{79}, 2193 (2001).

\bibitem{26} R. G. Mani et al., "Magnetoresistive response of a high
mobility 2DES under electromagnetic wave excitation," in the
\textit{Proc. of the 26th Intl. Conf. on the Phys. of Semicond.}
Edinburgh, Aug. 2002, IOP Conf. Ser. No. \textbf{171}, eds. A. R.
Long and J. H. Davies (IOP, Bristol, 2003) H112;
[cond-mat/0305507].

\bibitem{27} R. G. Mani et al., "Radiation induced oscillatory
magnetoresistance as a sensitive probe of the zero-field spin
splitting in high mobility GaAs/AlGaAs devices," \textit{Phys.
Rev. B} \textbf{69}, 193304 (2004).

\bibitem{28} V. I. Ryzhii, \textit{Fiz. Tverd. Tela (Leningrad)}
\textbf{11}, 2577 (1969) [\textit{Sov. Phys. - Sol. St.}
\textbf{11}, 2078-2080 (1970)].

\bibitem{29} J. C. Phillips, "Microscopic origin of collective exponentially
small resistance states," \textit{Sol. St. Comm. }\textbf{127},
233 (2003).

\bibitem{30} A. C. Durst, S. Sachdev, N. Read, and S. M. Girvin,
"Radiation-induced magnetoresistance oscillations in a 2D electron
gas," \textit{Phys. Rev. Lett.} \textbf{91}, 086803 (2003).

\bibitem{31} A. V. Andreev, I. L. Aleiner, and A. J. Millis, "Dynamical
symmetry breaking as the origin of the zero-dc-resistance state in
an ac-driven system," \textit{Phys. Rev. Lett.} \textbf{91},
056803 (2003).

\bibitem{32} P. W. Anderson and W. F. Brinkman, "New zero-resistance state
in heterojunctions: A dynamical effect," cond-mat/0302129.

\bibitem{33} J. Shi and X. C. Xie, "Radiation-induced zero-resistance state
and the photon-assisted transport," \textit{Phys. Rev. Lett.}
\textbf{91}, 086801 (2003).

\bibitem{34} A. A. Koulakov and M. E. Raikh, "Classical model for the negative dc conductivity of ac-driven
two-dimensional electrons near the cyclotron resonance,"
\textit{Phys. Rev. B} \textbf{68}, 115324 (2003).

\bibitem{35} F. S. Bergeret, B. Huckestein, and A. F. Volkov,
"Current-voltage characteristics and the zero-resistance state in
a two-dimensional electron gas," \textit{Phys. Rev. B} 67, 241303
(2003).

\bibitem{36} I. A. Dmitriev, A. D. Mirlin, and D. G. Polyakov, "Cyclotron-Resonance Harmonics in the ac Response of
a 2D Electron Gas with Smooth Disorder," \textit{Phys. Rev. Lett.}
\textbf{91}, 226802 (2003).

\bibitem{37} S. I. Dorozhkin, "Giant magnetoresistance oscillations caused by cyclotron resonance harmonics," \textit{JETP Lett.}, \textbf{77}, 577 (2003).

\bibitem{38} X. L. Lei and S. Y. Liu, "Radiation-induced magnetoresistance
oscillation in a two-dimensional electron gas in Faraday
geometry," \textit{Phys. Rev. Lett.} \textbf{91}, 226805 (2003).

\bibitem{39} V. Ryzhii and V. Vyurkov, "Absolute negative conductivity in
two-dimensional electron systems associated with acoustic
scattering stimulated by microwave radiation," \textit{Phys. Rev.
B} \textbf{68}, 165406 (2003).

\bibitem{40} D. H. Lee and J. M. Leinaas, "Role of interference in
millimeter-wave-driven dc transport in a two-dimensional electron
gas," \textit{Phys. Rev. B} \textbf{69}, 115336 (2004).

\bibitem{41} V. Ryzhii, "Microwave photoconductivity in two-dimensional
electron systems due to photon-assisted interaction of electrons
with leaky interface phonons," \textit{Phys. Rev. B} \textbf{68},
193402 (2003).

\bibitem{42} M. G. Vavilov and I. L. Aleiner, "Magnetotransport in a
two-dimensional electron gas at large filling factors,"
\textit{Phys. Rev. B} \textbf{69}, 035303 (2004).

\bibitem{43} R. Klesse and F. Merz, "Residual resistance in two-dimensional microwave driven systems," preprint cond-mat/0305492.

\bibitem{44} A. F. Volkov and V. V. Pavlovskii, "Residual resistance in a two-dimensional electron system:
A phenomenological approach,"  \textit{Phys. Rev. B }\textbf{69},
125305 (2004).

\bibitem{45} V. Ryzhii and A. Satou, "Electric-Field Breakdown of Absolute Negative Conductivity and Supersonic Streams
in Two-Dimensional Electron Systems with Zero
Resistance/Conductance States," \textit{J. Phys. Soc. Jpn.}
\textbf{72}, 2718 (2003).

\bibitem{46} R. G. Mani et al., "Radiation induced zero-resistance
states in GaAs/AlGaAs heterostructures: Voltage-current
characteristics and intensity dependence at the resistance
minima," preprint cond-mat/0306388.

\bibitem{47} R. G. Mani, "Zero-resistance states induced by
electromagnetic-wave excitation in GaAs/AlGaAs heterostructures,"
\textit{Physica E }\textbf{22}, 1 (2004) .

\bibitem{48} R. G. Mani et al., "Demonstration of a 1/4-Cycle Phase Shift in the Radiation-Induced
Oscillatory Magnetoresistance in GaAs/AlGaAs Devices,"
\textit{Phys. Rev. Lett. }\textbf{92}, 146801 (2004).

\bibitem{49} R. G. Mani et al., "Radiation-induced oscillatory Hall effect in
high-mobility $GaAs/Al_{x}Ga_{1-x}As$ devices," \textit{Phys. Rev.
B} \textbf{69}, 161306 (2004).

\bibitem{50} S. A. Studenikin, M. Potemski, P.T. Coleridge, A. Sachrajda, and Z.R. Wasilewski
"Microwave radiation induced magneto-oscillations in the
longitudinal and transverse resistance of a two dimensional
electron gas", \textit{Sol. St. Comm.} \textbf{129}, 341 (2004).

\bibitem{51} A. W. Overhauser, "Polarization of nuclei in metals," \textit{Phys. Rev.}
\textbf{92}, 411 (1953)

\bibitem{52} J. M. Taylor, C. Marcus, and M. D. Lukin, "Long-lived memory
for mesoscopic quantum bits," \textit{Phys. Rev. Lett.
}\textbf{90}, 206803 (2003).

\bibitem{53} Y. A. Bychkov and E. I. Rashba, "Oscillatory effects and
magnetic suscepibility of carriers in inversion layers,"
\textit{J. Phys. C. }\textbf{17}, 6039-6045 (1984).

\bibitem{54} G. Dresselhaus, "Spin orbit coupling effects in zinc blende
structures," \textit{Phys. Rev. }\textbf{100}, 580 (1955).

\bibitem{55} G. Lommer, F. Malcher, and U. R\"{o}ssler, "Spin splitting in
semiconductor heterostructures for B $\rightarrow$ 0,"
\textit{Phys. Rev. Lett.} \textbf{60}, 728-731 (1988).

\bibitem{56} R. Eppenga and M. F. H. Schuurmans, "Effect of bulk inversion
asymmetry on [001], [110], and [111] GaAs/AlAs quantum wells,"
\textit{Phys. Rev. B. }\textbf{37}, 10923-10926 (1988).

\bibitem{57} E. A. De Andrada e Silva, G. C. La Rocca, and F. Bassani,
"Spin-split subbands and magneto-oscillation in III-V saymmetric
heterostructures," \textit{Phys. Rev. B }\textbf{50}, 8523-8533
(1994).

\bibitem{58} B. Das et al., "Evidence for spin splitting in InGaAs/InAlAs
heterostructures as B $\rightarrow$ 0,"  \textit{Phys. Rev. B}
\textbf{39}, 1411 - 1414 (1989).

\bibitem{59} P. Ramvall, B. Kowalski, and P. Omling, "Zero-magnetic fields
spin splittings in AlGaAs/GaAs heterojunctions," \textit{Phys.
Rev. B }\textbf{55}, 7160-7164 (1997).

\bibitem{60} A. C. H. Rowe, J. Nehls, R. A. Stradling, and R. S. Ferguson,
"Origin of beat patterns in the quantum magnetoresistance of gated
InAs/GaSb and InAs/AlSb quantum wells," \textit{Phys. Rev. B}
\textbf{63}, 201307 (2001).

\bibitem{61} T. H. Sander et al., "Determination of the phase of
magneto-intersubband scattering oscillations in heterojunctions
and quantum wells," \textit{Phys. Rev. B }\textbf{58}, 13856-13862
(1998).

\bibitem{62} B. Jusserand, D. Richards, G. Allan, C. Priester, and B.
Etienne, "Spin orientation at semiconductor heterointerfaces,"
\textit{Phys. Rev. B} \textbf{51}, 4707-4710 (1995).

\bibitem{63} J. B. Miller et al., "Gate-controlled spin-orbit quantum
 interference effects in lateral transport," \textit{Phys. Rev. Lett. }\textbf{90}, 076807
(2003).

\bibitem{64} S. A. Tarasenko and N. S. Averkiev,
"Interference of spin splittings in magneto-oscillation phenomena
in two-dimensional systems" \textit{JETP Letters}, \textbf{75},
669-672, 2002.

\bibitem{65} S. Saykin, M. Shen, M. -C. Cheng, and V. Privman,
"Semiclassical Monte-Carlo mode for in-plane transport of
spin-polarized electrons in III-V heterostructures,"  \textit{J.
Appl. Phys.} \textbf{94}, 1769-1776 (2003); S. Saykin, "A Drift
diffusion model for spin polarized transport in a non-degenerate
2DEG controlled by spin-orbit interaction," \textit{J. Phys.
Condens. Matter} \textbf{16}, 5071-5081 (2004).

\bibitem{66} Y. V. Pershin, J. A. Nesteroff, and V. Privman,
"Effect of spin-orbit interaction and in-plane magnetic field on
the conductance of a quasi-one-dimensional system," Phys. Rev. B
\textbf{69}, 121306 (2004); Y. V. Pershin, "Drift diffusion
approach to spin-polarized transport," Physica E \textbf{23},
226-231 (2004).



\end{thebibliography}

\end{document}